\documentclass[twocolumn,prb]{revtex4-2}
\usepackage{graphicx,amssymb,soul,color,amsmath,bm,dcolumn}
\usepackage{epstopdf,hyperref}
\usepackage{braket,dsfont,nicefrac,bbold}
\usepackage[mathcal]{euscript}
\usepackage[normalem]{ulem}
\usepackage[utf8]{inputenc}
\usepackage{float}

\hypersetup{colorlinks=true,citecolor=blue,linkcolor=magenta}

\sethlcolor{yellow}
\allowdisplaybreaks

\begin{document}

\title{Phase diagram of spin-3/2 fermions in one-dimensional optical lattices}

\author{Samuel Milner and Adrian Feiguin}
\affiliation{Physics Department, Northeastern University, Boston, MA 02115, USA}

\begin{abstract}
We present a density matrix renormalization group (DMRG) study of a generalized Hubbard chain describing effective spin $S = 3/2$ fermions in an optical lattice.  We determine the full phase diagram  for the $SU(4)$ symmetric case, and in the presence of ``single-ion anisotropy'' in terms of density and polarization. We investigate the stability and competition between different orders, such as ``quintet'' Fulde-Ferrell-Larkin-Ovchinnikov (FFLO) pairing, trion and quartet formation, and spin and atomic density waves. Notably, near half-filling, single-ion anisotropy stabilizes a correlated phase that can be understood in terms of a generalized $S=2$ {\it bosonic} $t-J$ chain. 
\end{abstract}

\date{\today}
\maketitle

\section{Introduction}
   The study of ultracold gases offers a gateway to rich and unique physical phenomena compared to those seen in solid state physics\cite{bloch_review,Lewenstein2007}.  The interaction between electrons and nuclei generates a high hyperfine spin $F$, which allows for the uncovering of unconventional correlated phases \cite{Wu2006,Capponi2016Phases}. These  states with $SU(N)$ symmetry seen in the rare alkali earth atoms such as Be$^{9}$, Ba$^{137}$, Ba$^{135}$, Li$^{6}$ and Cs$^{137}$, with $N$ being proportional to nuclear spin ($N=2F+1$) \cite{Honerkamp2004,Wu2003Exact,Gorshkov2010,Manmana2011,Capponi2016Phases}. 
   
   Interest in high-spin ensembles has extended to condensed matter, where the effects of strong spin-orbit coupling in  materials such as $RT$Bi ($R$=rare earth, $T$=Pt or Pd) that can be classified as ``half-Heusler'' semimetals can lead to exotic paired states beyond the BCS paradigm\cite{Boettcher2016,Brydon2016,Kim2018a}. The spin orbit coupling between $S = 1/2$ spins and the $L = 1$ $p$-band in the Bi atoms for these compounds results in $J =3/2$ electronic states, which allows for the generation of higher angular momentum Cooper pairs such as \textit{quintet} or even \textit{septet} paired states. 
   
   In this work we will not consider Hamiltonians or physical models that include spin-orbit coupling, rather we shall focus on a Hubbard-like chain with a spin species corresponding to a specific hyperfine spin value $F$.  For the case of $F = 3/2$, choosing equal interaction strengths in the $S=0$ and $S=2$ channels generates an $SU(4)$ symmetry in the $s$-wave scattering channel, a regime favorable for experiments concerning ultracold fermionic gases. The $S =1$ and $S = 3$ channels for $F = 3/2$ are forbidden for fermions, a consequence of Pauli's exclusion principle\cite{Wu2003Exact,Wu2005Competing,Tu2006, Capponi2007Confinement,Jiang2015,Brydon2016,Capponi2016Phases}.
   
   Previous research for the case of two component spinor fermions in one spatial dimension revealed the presence of a Fulde-Ferrell-Larkin-Ovchinnikov (FFLO) state\cite{FF,LO} for high interaction strength \cite{Feiguin2007Pairing,Batrouni2009Exact,Feiguin2009Pair,Feiguin2009Spectral,Heidrich-Meisner2010BCS,Heidrich-Meisner2010Phase,Dalmonte2012}. The FFLO state is characterized by the formation of Cooper pairs with finite momenta $K$ compared to the non-zero momenta seen in conventional BCS theory \cite{FF,LO}.

   For the case of atomic systems with hyperfine spin $F > 1/2$, FFLO-like states have been theorized, as well as the existence of ``quintet'' pairing for different spin channels, and even quartet pairing\cite{Lecheminant2005Confinement,Lecheminant2008,Roux2009Spin,Barcza2012,Jiang2015,Szirmai2017}. 
   Moreover, for $F=3/2$ they may realize hidden symmetries \cite{Wu2006,Wu2005Competing,Capponi2016Phases} and an exactly solvable (integrable) $SO(4)$-symmetric point\cite{Jiang2015}. Previous numerical work on such spin-3/2 chains revealed the presence of emergent superfluidity and FFLO-like pairing\cite{Barcza2012}. 
   Given the high precision and accuracy of numerical techniques in 1-D such as density matrix renormalization group (DMRG)\cite{White1992,White1993}, a multi-spin component single chain offers an ideal scenario to investigate the rich physics of these systems. 
   In this work we complement previous studies by calculating the full phase diagram of a chain of spin-3/2 interacting fermions in one-dimensional optical lattices as a function of chemical potential and magnetic field in different interacting regimes. The paper is organized as follows: In Sec.\ref{sec:model} we introduce the model Hamiltonian, the different order parameters associated to it, and computational details; in Sec.\ref{sec:results} we present and discuss results for the phase diagrams as a function of chemical potential and magnetic field; finally, we close with a summary and conclusion.

   \section{Model and method}\label{sec:model}

We consider the simplest possible model describing fermions with hyperfine angular momentum $F=3/2$ in a one-dimensional optical lattice. Introducing a contact interaction in the s-wave scattering channel (From now on, we will refer to the angular momentum as ``spin'' with the symbol $S$), it is written as: 
\begin{eqnarray}
H &=& -J\sum\limits_{\langle ij \rangle,\alpha } \left(
c_{i,\alpha}^{\dagger}c_{j,\alpha }
+\mathrm{H.c.} \right) + \nonumber \\
&+& U_0\sum\limits_{i}P^\dagger_{00,i} P_{00,i} + U_2 \sum\limits_{i,m} P^\dagger_{2m,i}P_{2m,i} \\
&-&h\sum_i S_i^z -\mu\sum_i N_i, \nonumber
\label{hami0}
\end{eqnarray} 
   with hopping matrix elements $J$ between nearest neighbors $\langle ij\rangle$. The operators $c^\dagger_{i,\alpha}$ refer to fermionic creation operators acting on a site $i$ for a given $S^z$ projection of spin $\alpha = \pm 1/2, \pm 3/2$. The interaction parameters are proportional to the scattering lengths in the various spin $S$ channels, and thus can be tuned in accordance to the ``depth'' or potential of an optical trap. We have also added an external magnetic field $h$ and a chemical potential $\mu$. The operators $P^\dagger_{Sm}$ create a pair with total spin $S$ 
 and projection $m$  as\cite{Barcza2012,Tu2006,Capponi2007Confinement}
\begin{equation}
    P^\dagger_{S,m,i}=\sum_{\alpha,\alpha'} \langle\frac{3}{2},\frac{3}{2};\alpha,\alpha'|S,m\rangle c^\dagger_{i,\alpha}c^\dagger_{i,\alpha'},
\label{pair}
\end{equation}
with $(S,m)=\{(0,0);(2,\pm 2);(2,\pm 1;)(2,0)\}$. In this expression we introduced the Clebsch-Gordan coefficients for addition of two spin-3/2 angular momenta. The singlet and quintet paired states 
 (those in the $S=2$ sector) for our spin-3/2 chain are written accordingly in Table 1. 
 
 \begin{table}[b]
 \caption{\label{tab:table1} Singlet and Quintet Pairing terms}
\begin{ruledtabular}
\begin{tabular}{lcdr} 
 \hline
 Pair Term&
2nd. Quantization &
{\it (S,m)} &
\\
\colrule
 \hline\hline

$P^\dagger_{2,2}$ & $c^\dagger_{3/2}c^\dagger_{1/2}$ & (2,2) \\ [0.15cm]
 
 $P^\dagger_{2,1}$  &   $c^\dagger_{3/2}c^\dagger_{-1/2}$ & (2,1)\\ [0.15cm]
  
  $P^\dagger_{2,-1}$  &   $c^\dagger_{-3/2}c^\dagger_{1/2}$ &(2,-1) \\ [0.15cm]
   
  $P^\dagger_{2,-2}$  & $c^\dagger_{-3/2}c^\dagger_{-1/2}$ & (2,-2)\\[0.15cm]

 $P^\dagger_{2,0}$ & $\frac{1}{\sqrt{2}}[c^\dagger_{3/2}c^\dagger_{-3/2}+c^\dagger_{1/2}c^\dagger_{-1/2}]$ & (2,0)  \\
 \hline
 
  $P^\dagger_{0,0}$ & $\frac{1}{\sqrt{2}}[c^\dagger_{3/2}c^\dagger_{-3/2}-c^\dagger_{1/2}c^\dagger_{-1/2}]$ & (0,0)  \\
 
 \hline
 \end{tabular}

 \end{ruledtabular}
 \end{table}

  It is possible to express the interactions in terms of the density and spin operators as\cite{Tu2006}:
\begin{eqnarray}
H &=& -J\sum\limits_{\langle ij \rangle,\alpha } \left(
c_{i,\alpha}^{\dagger}c_{j,\alpha }
+\mathrm{H.c.} \right) + 
\frac{U}{2}\sum\limits_{i}N_{i}^2  + \frac{V}{2}\sum\limits_{i}\mathbf{S}_{i}^2+ \nonumber \\
&-&h\sum_i S_i^z -\mu\sum_i N_i,
\label{hami}
    \end{eqnarray}  
 where $N_i=\sum_\alpha n_{\alpha,i}$ and $\mathbf{S}_i=\sum_{\alpha\beta}c^\dagger_{\alpha,i}\mathbf{S}_{\alpha\beta}c_{\beta,i}$ with the spin represented by spin-3/2 matrices. In deriving Eq.~(\ref{hami}) we have used the identities:
 \begin{eqnarray}
 \mathbf{S}^2_i-\frac{15}{4}N_i &=& -\frac{3}{2}\sum_{m}P^\dagger_{2m,i}P_{2m,i}-\frac{15}{4}P^\dagger_{00,i}P_{00,i} \nonumber \\
N^2_i-2N_i&=&\sum_{m}P^\dagger_{2m,i}P_{2m,i}+P^\dagger_{00,i}P_{00,i}.
 \end{eqnarray}
Accordingly, the constants in the two models are related as: $U_0=\frac{U}{2}-\frac{15}{8}V; U_2=\frac{U}{2}-\frac{3}{4}V$.

   In the absence of the ``single-ion anisotropy'', $V=0$, the Hamiltonian reduces to the $SU(4)$ Hubbard model \cite{Honerkamp2004}, invariant under local rotations of the 4 flavors with different $\alpha$. We notice that the Hamiltonian conserves total ``charge'' or number of particles $N$ and spin projection $S^z$ (we point out that ``charge'' is not rigorously an appropriate label to refer to neutral atoms, but it is used interchangeably in analogy to condensed matter systems). 

\begin{figure}
    \includegraphics[width=\linewidth]{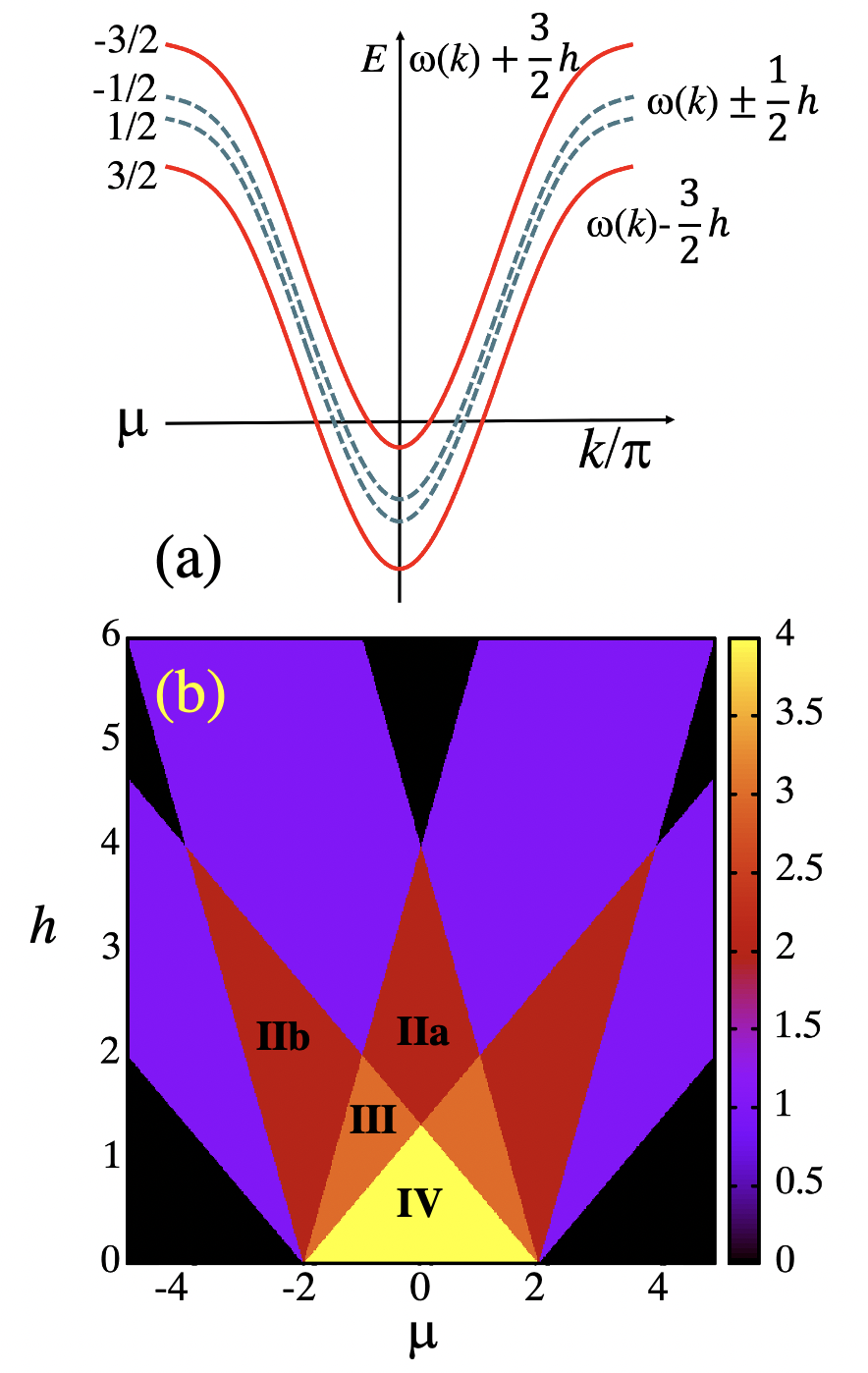}
    \caption{Non-interacting regimes: (a) band splitting under an external magnetic field; the chemical potential can intersect up to four bands at a time, with $\omega(k)=-2J\cos{k}$. (b) Color density plot identifying the regions according to the number of the number of partially filled bands. Region IV corresponds to four partially filled bands; III indicates three partially filled bands, with spins $(\pm 1/2,3/2)$; IIa and IIb correspond to two partially filled bands with spins $(3/2,1/2)$ and $\pm 1/2$, respectively.  }
    \label{fig:free_fermions}
     \end{figure}

    It is instructive to consider the non-interacting limit as a guide. In the absence of a magnetic field the four bands corresponding to different values of $\alpha$ are degenerate. The magnetic field will split them according to their $S^z$ quantum number $\alpha={-3/2,-1/2,1/2,3/2}$ and, correspondingly, the chemical potential will intersect them in (up to) eight Fermi points $\pm k_{F\alpha}$, depending on the filling fraction/density of fermions (see Fig.\ref{fig:free_fermions})(a). As the chemical potential increases, the bands will start filling up one by one, starting with the $\alpha=3/2$ one, followed by $\alpha=1/2$ and so on. According to the FFLO prediction, in the presence of attractive ($U < 0$) interaction, we should expect the fermions to pair with center of mass momentum $K=0$ (conventional Cooper pairs) or finite center-of-mass momentum $K=k_{F\alpha}-k_{F\alpha'}$. For strong enough attraction, this term may induce the formation of quartets instead of pairs. 

The addition of the single-ion anisotropy will favor the formation of pairs with total spin $S=2(0)$ for negative (positive) values of $V$, but will compete with(or favor) the formation of quartets, that have $S=0$.

    We determine the stability phase diagrams as a function of magnetic field $h$ and chemical potential $\mu$ for various interaction strengths. The grand canonical ensemble may reveal incompressible phases, and that some phases may not be energetically stable, with the system ``phase separating'' into domains. 

    We can characterize the different phases by calculating correlation functions for ``quintet'' and singlet pairing in the $(S,m)$ channels. We also consider the possibility of quartet pairing through the operator:
    \begin{equation}
    Q^\dagger_i = c^\dagger_{i,3/2}c^\dagger_{i,-3/2}c^\dagger_{i,1/2}c^\dagger_{i,-1/2},
    \label{quartet}
    \end{equation}
    and the formation of a trion quasi-condensate is explored through the operator 
    \begin{equation}
    T^\dagger_i = c^\dagger_{i,3/2}c^\dagger_{i,1/2}c^\dagger_{i,-1/2}.
    \label{trion}
    \end{equation}
    
  Correlation functions for various pairing states, as well as the quartet state, are also obtained in momentum space as 
   \begin{equation}\label{paireq}
   P_{S,m}(k) = \sum_{r,s} e^{ik(r-s)} \langle P^\dagger_{S,m,r}P_{S,m,s} \rangle.
   \end{equation}
   Similar expressions are derived for the quartet correlator, that we refer-to as $Q(k)$, longitudinal spin structure factor $S^z(k)$, and density structure factor $D(k)$.

   We point out that, for $V=0$, the Hamiltonian conserves particle number on each channel $\alpha$ and the system does not break $U(1)$ symmetry. Hence, the expectation value of correlators such as $\langle c^\dagger_{3/2}c_{1/2}\rangle =0 $, always, and the pair-pair correlations in the $m=0$ channels become indistinguishable:
   \begin{equation}
       \langle P^\dagger_{S,0}P_{S,0}\rangle=\frac{1}{2}\left(\langle \Delta^\dagger_{3/2}\Delta_{3/2}\rangle+\langle\Delta^\dagger_{1/2}\Delta_{1/2}\rangle\right),
   \end{equation}
 where we introduce the operators $\Delta^\dagger_{\alpha}=c^\dagger_{\alpha}c^\dagger_{-\alpha}$, with $\alpha=3/2,1/2$. This is no longer the case in the presence of non-zero $V$, since the terms $P^\dagger_{S,0}P_{S,0}$ in the Hamiltonian contains cross terms that will exchange pairs of particles between the $\alpha=\pm 3/2$ and $\alpha'=\pm 1/2$ channels.

 In the presence of a magnetic field, time-reversal symmetry will be broken, and makes more sense to think of pairing channels with well defined $\alpha,\alpha'$, instead of total spin $S$. Therefore, at finite fields, we shall study the $\Delta_\alpha$ correlations instead of those for $P_{2,0}$ and $P_{0,0}$ .


     \begin{figure}
    \includegraphics[width=\linewidth]{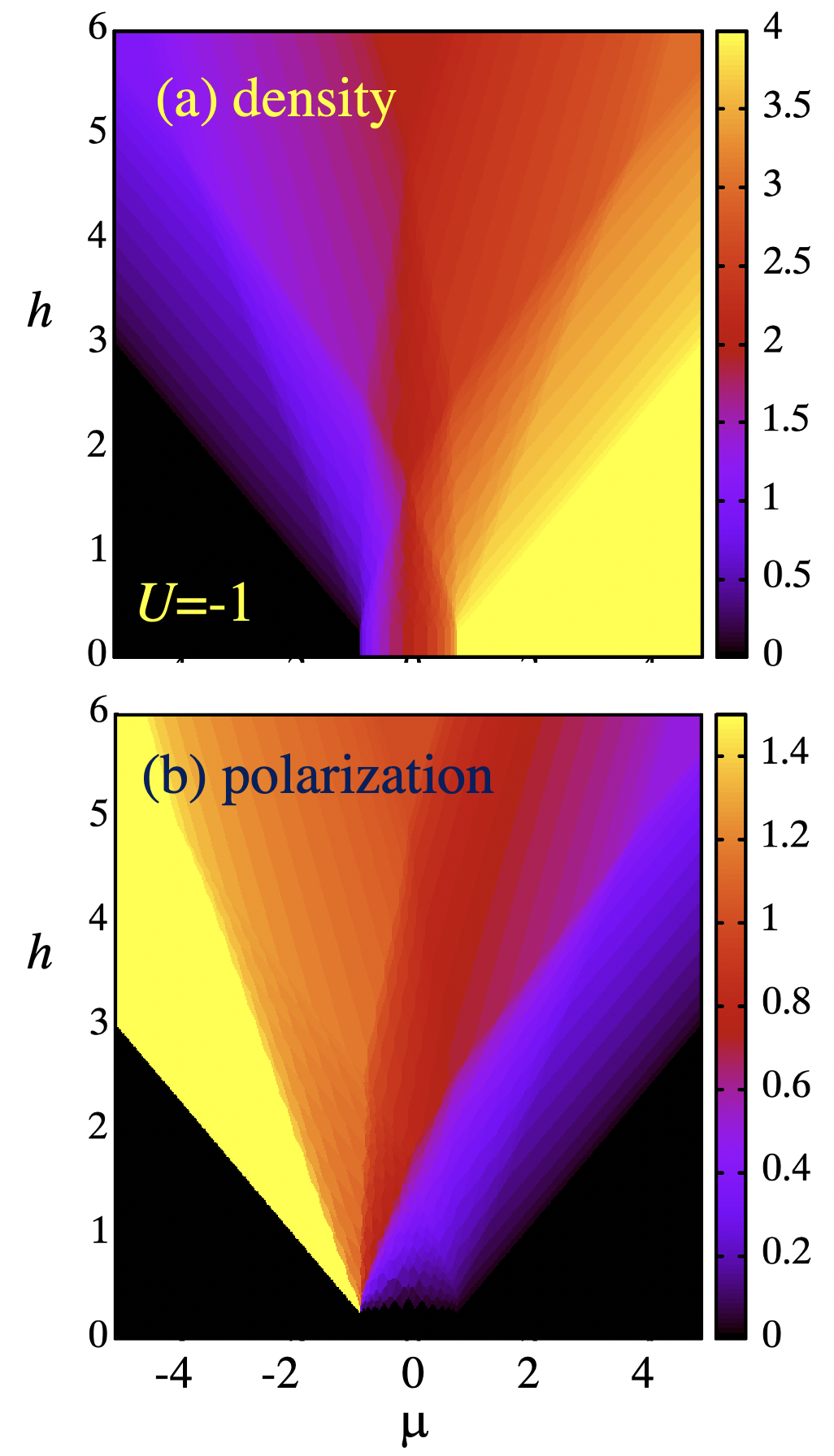}
    \caption{(a) Density $N/L$ and (b) polarization $S^z/N$ as a function of field and chemical potential for $U=-1; V=0$ obtained for a chain with $L=16$ sites using DMRG. }
    \label{fig:phases_U-1}
     \end{figure}

\section{Results}\label{sec:results}
\subsection{$SU(4)$ case}

\subsubsection{Phase diagram}

In order to gain intuition on the problem and predict possible pairing scenarios, we first analyze the phase diagram of the non-interacting chain as a function of field and chemical potential, as shown in Fig.\ref{fig:free_fermions}. In Fig.\ref{fig:free_fermions}(a) we depict the non-interacting bands with the broken degeneracy due to the magnetic field. As a consequence, the chemical potential may cut one, two, three, or four bands simultaneously, depending on the the magnitude of $h$. As the chemical potential increases, bands may become filled, one at a time. In Fig.\ref{fig:free_fermions}(b) we show the different regimes, with the color density indicating the number of bands that are partially filled. At low fields, low densities, four bands will be partially filled; as the field increases, we transition to regimes where the band corresponding to $\alpha=3/2$ is filled, and the other three are partially filled, and so on. At large magnetic fields, the four bands do not overlap in energy and only one band is partially filled at a time, with the other three either full, or empty. In these cases we expect the physics to be the same as spinless fermions, and uninteresting. 

\begin{figure}
    \centering
    \includegraphics[width=\linewidth]{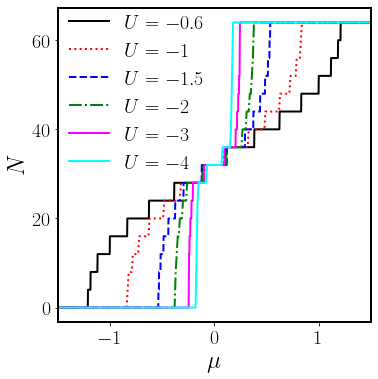}
    \caption{Particle number as a function of chemical potential for a chain of length $L=16$ and several values of $U$, obtained by means of the Maxwell construction (see text for details). The chemical potential is shifted by $2U$ to center at the particle-hole symmetric point.}
    \label{fig:nvsmu}
     \end{figure}

Using the Hamiltonian in Eq.~\ref{hami} we can generate the phase diagram for the interacting case by first calculating ground state energies in the canonical ensemble $E_0(N,S^z)$ using the DMRG method for open boundary conditions with $L$ = 16 sites, for fixed number of particles $N$ and spin $S^z$. We use enough DMRG states to guarantee a convergence in the energy to 7 decimal places. By means of a Maxwell construction, we obtain the stability of each state $(N,S^z)$ as a function of $\mu$ and $h$. This process consists of fixing the values of $(\mu,h)$ and searching for the pair $(N,S^z)$ that minimizes $E(\mu,h)=E_0(N,S^z)-\mu N -h S^z$. The solution to this problem yields the occupation and polarization as a function of $(\mu,h)$.
In Figs.\ref{fig:phases_U-1} we show results for $U=-1$. We point out that the chemical potential was shifted by $2U$ in order to make evident the particle-hole symmetry of the model. This also makes it easier to contrast and correlate the results with those in the non-interacting case, Fig. \ref{fig:free_fermions}. There is a clear one-to-one correspondence between the different ``diamonds'' and ``triangles'' in both cases, allowing us to identify different regimes according to the number of gapless ``charge modes''. In the following we focus on the hole-doped regime, for negative $\mu<0$.

         \begin{figure}
    \centering
    \includegraphics[width=\linewidth]{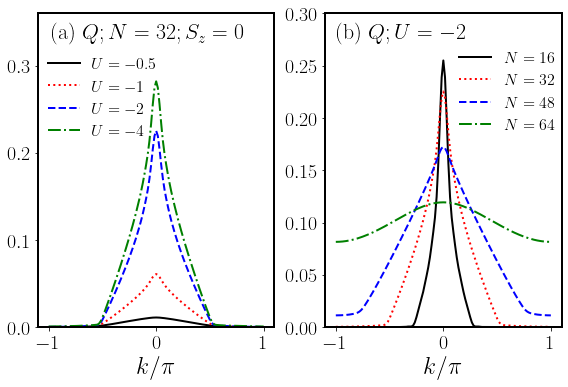}
    \caption{Quartet correlations in momentum space for (a) quarter-filling ($L=N=32$, $S^z=0$) and different values of $U, V=0$; (b) for $U = -2, V=0$ and different values of $N$.}
    \label{fig:pairQ}
     \end{figure}

\subsubsection{Unpolarized states}

    We begin by first identifying a region that is not in the non-interacting phase diagram: At all densities we encounter an unpolarized phase -- seen in black in Figs.\ref{fig:phases_U-1}(b) -- that extends to small but finite magnetic fields because, as we will establish, the system is spin gapped.  

     We examine the properties of the ground state in this regime for the $SU(4)$ sector ($V$ = 0) and $U < 0$. In Fig. \ref{fig:nvsmu} we show the density as a function of chemical potential at zero polarization, $S^z=0$, for a chain of length $L=16$ and several values of $U$. We clearly notice as the density varies in steps of 4 for sufficiently small $|U|$. This indicates that it is energetically more favorable to add particles in quartets. 
     
    The presence of quartet pairing is a known phenomena for $S$ = 3/2 fermionic spin chains. 
    In the thermodynamic limit, it costs energy to break a quartet, but not to add one (same as for doublons in the attractive Hubbard model) and the system therefore is spin gapped, but density(charge) gapless\cite{Roux2009Spin}. As $U$ becomes increasingly more negative, we observe a trend toward phase separation, as  previously noted in other studies \cite{Roux2009Spin,Szirmai2017}. When this occurs, quartets will form domains or ``bubbles''. These configurations are not thermodynamically stable, manifesting into abrupt jumps in the $N$ vs. $\mu$ curve. Therefore, in the following we concentrate in the regime with moderately small $|U|$.

     \begin{figure}
    \centering
    \includegraphics[width=\linewidth]{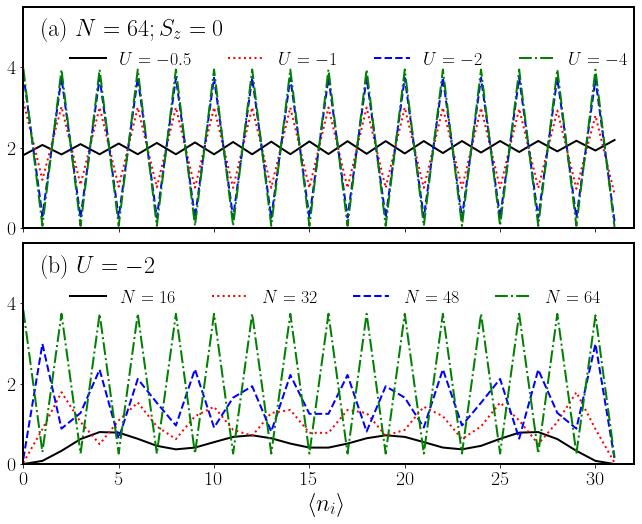}
    \caption{Local density in real space $\langle N_i\rangle$ (a) half-filling ($L=32$; $N=64$; $S^z=0$) and different values of $U, V=0$; (b) for $U = -2, V=0$ and different values of $N$.}
    \label{fig:density}
     \end{figure}


    In Fig.\ref{fig:pairQ} we show results for the  quartet correlation function in momentum space for a chain of length $L = 32$ at zero polarization and for several densities and values of $U$ ($V$ = 0). The sharp singular cusp at smaller densities and larger $U$ indicates the existence of a quasi-condensate for densities $N \le L$ (quarter filling). On the other hand, the triangular shape for larger densities $N=48$ corresponds to correlations that decay as $\sim 1/x$, while the flat one for $N=64$ indicates that quartets are localized and uncorrelated. In all these cases, the pair correlations are essentially featureless (not shown), implying that pairing  is not favored.

    However, as was pointed out in Refs.\onlinecite{Roux2009Spin,Szirmai2017}, there is an additional competing instability toward an ``atomic density wave'' (ADW) -- notice again that referring to ``charge'' is not appropriate in the context of neutral atoms. This order is described by a spatial modulation of the particle density. In the presence of open boundary conditions, these modulations are  automatically pinned as clearly observed in Fig.\ref{fig:density}. We find that at half-filling the system is in a robust ADW phase in the full range of values of $U$ (see Fig.\ref{fig:density}(a)). However, as seen in panel Fig.\ref{fig:density}(b), the ADW no longer dominates at lower densities, and the system favors a quartet quasi-condensate, consistent with the findings in the previous figure, Fig.\ref{fig:pairQ}. 

    \subsubsection{Paired states at finite magnetic Field}

In this section we investigate other exotic paired states by introducing a finite polarization or spin imbalance. 
 
     We proceed by examining correlations in momentum space for various values of external magnetic field $h$. We focus first on region IV. As seen in the illustration of Fig. \ref{fig:free_fermions}(a), the system has 8 gapless modes with Fermi momenta $\pm k_{F\alpha}$. According to the FFLO prediction, pairs can form between channels $\alpha,\alpha'$ with different symmetry and with finite center of mass momentum $K=k_{F\alpha}-k_{F\alpha'}$. 
     
\begin{figure}
    \includegraphics[width=\linewidth]{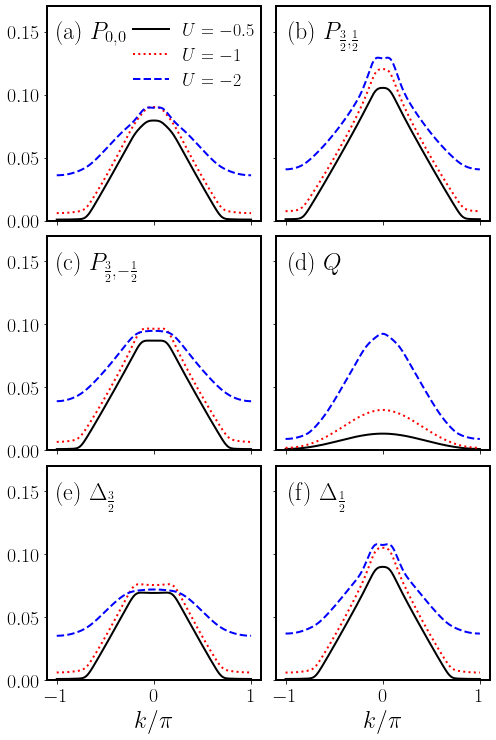}
    \caption{Pair-pair and quartet correlations for a chain of $L=32$ sites, $N=44, 2S^z=20$,  corresponding to a point inside region IV, with 8 gapless modes. The particle numbers for each channel are $N_{3/2}=14$; $N_{-3/2}=8$; $N_{1/2}=12$; $N_{-1/2}=10$. We show results for different values of $U=-0.5,-1,-2$.}
    \label{fig:pair4}
     \end{figure}

     In Fig.\ref{fig:pair4} we pick an arbitrary point with $N=44$ and $2S^z=20$ on a chain with $L=32$ sites, well within this region. The particle numbers for each channel are $N_{3/2}=14$; $N_{-3/2}=8$; $N_{1/2}=12$; $N_{-1/2}=10$. We see features that are not conclusive of pairing. While kinks at finite momentum are observed for some correlations, there is no clear singular behavior that can be associated to a quasi-condensate. With increasing $|U|$, the quartet correlations grow as the pair-pair correlations become more featureless. However, the smooth Gaussian-like profile in Fig.\ref{fig:pair4}(d) indicates exponentially decaying correlations and absence of quasi-long range order. In addition, the density of particles is uniformly distributed along the chain, without the characteristic modulations of the ADW phase (not shown). Hence, we conclude that this phase is simply a featureless metallic phase.

\begin{figure}
    \includegraphics[width=\linewidth]{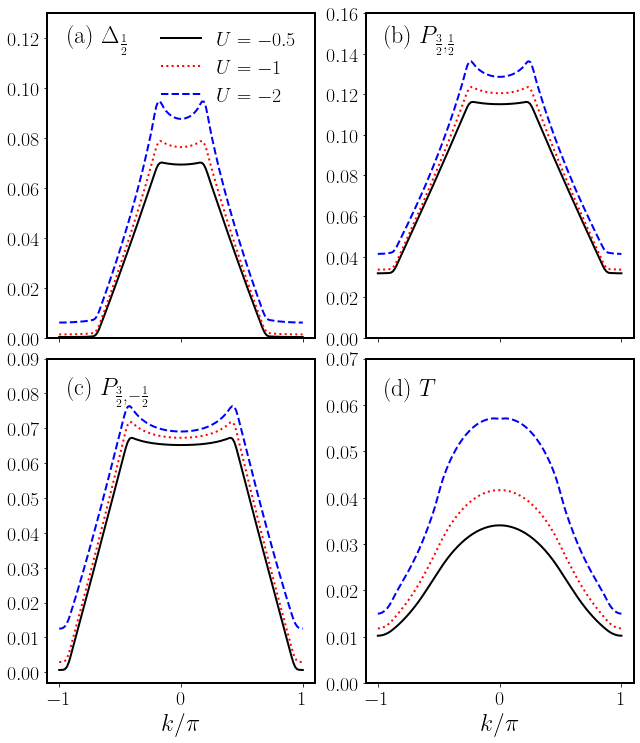}
    \caption{Pair-pair and trion correlations for a chain of $L=32$ sites, $N=44, 2S^z=72$,  corresponding to a point inside region III, with 3 gapless modes. We show results for different values of $U=-0.5,-1,-2$. The particle numbers for each channel are $N_{3/2}=22$; $N_{-3/2}=0$; $N_{1/2}=14$; $N_{-1/2}=8$. }
    \label{fig:pair3}
     \end{figure}

We now shift our attention on phase III with six gapless modes, as shown in Fig.\ref{fig:pair3}. In this regime, the band with spin $\alpha=-3/2$ is completely empty, and the other three are partially filled. Hence, we can also expect, besides pairing, the possibility of trion formation for sufficiently large $|U|$. Trions are fermionic quasiparticles, and the momentum distribution function is expected to have a singular behavior at some characteristic Fermi vector. However, what we observe instead is an increased pairing trend in all channels as inferred from the cusp-like features emerging at finite momentum. This suggests the coexistence of competing pairing instabilities.

\begin{figure}
    \includegraphics[width=\linewidth]{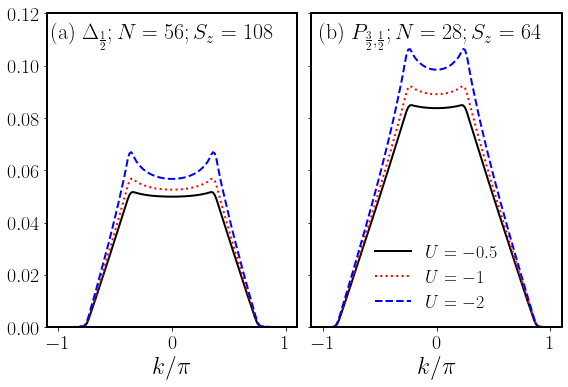}
    \caption{Pair-pair correlations for a chain of $L=32$ sites. We show results for (a) $N=56, 2S^z=108$ and (b), $N=28, 2S^z=64$, corresponding to points inside region IIa and IIb, respectively. The particle numbers per channel are $N_{3/2}=32$; $N_{-3/2}=0$; $N_{1/2}=18$; $N_{-1/2}=6$ and $N_{3/2}=18$; $N_{-3/2}=0$; $N_{1/2}=10$; $N_{-1/2}=0$.}
    \label{fig:pairw}
     \end{figure}

The physics of regions IIa and IIb is more conventional and easier to understand. In these cases, only two bands are partially filled, corresponding to $(\alpha=1/2,\alpha'=3/2)$ and $(\alpha=1/2,\alpha'=-1/2)$, respectively. In both cases, there is only one possible pairing channel, and in Fig.\ref{fig:pairw} we observe correlations reminiscent of those found in spin $S=1/2$ chains\cite{Feiguin2007Pairing,Batrouni2009Exact,Feiguin2009Pair,Feiguin2009Spectral,Heidrich-Meisner2010BCS,Heidrich-Meisner2010Phase,Dalmonte2012}.

\subsection{Beyond $SU(4)$ symmetries}
     \begin{figure}
    \includegraphics[width=\linewidth]{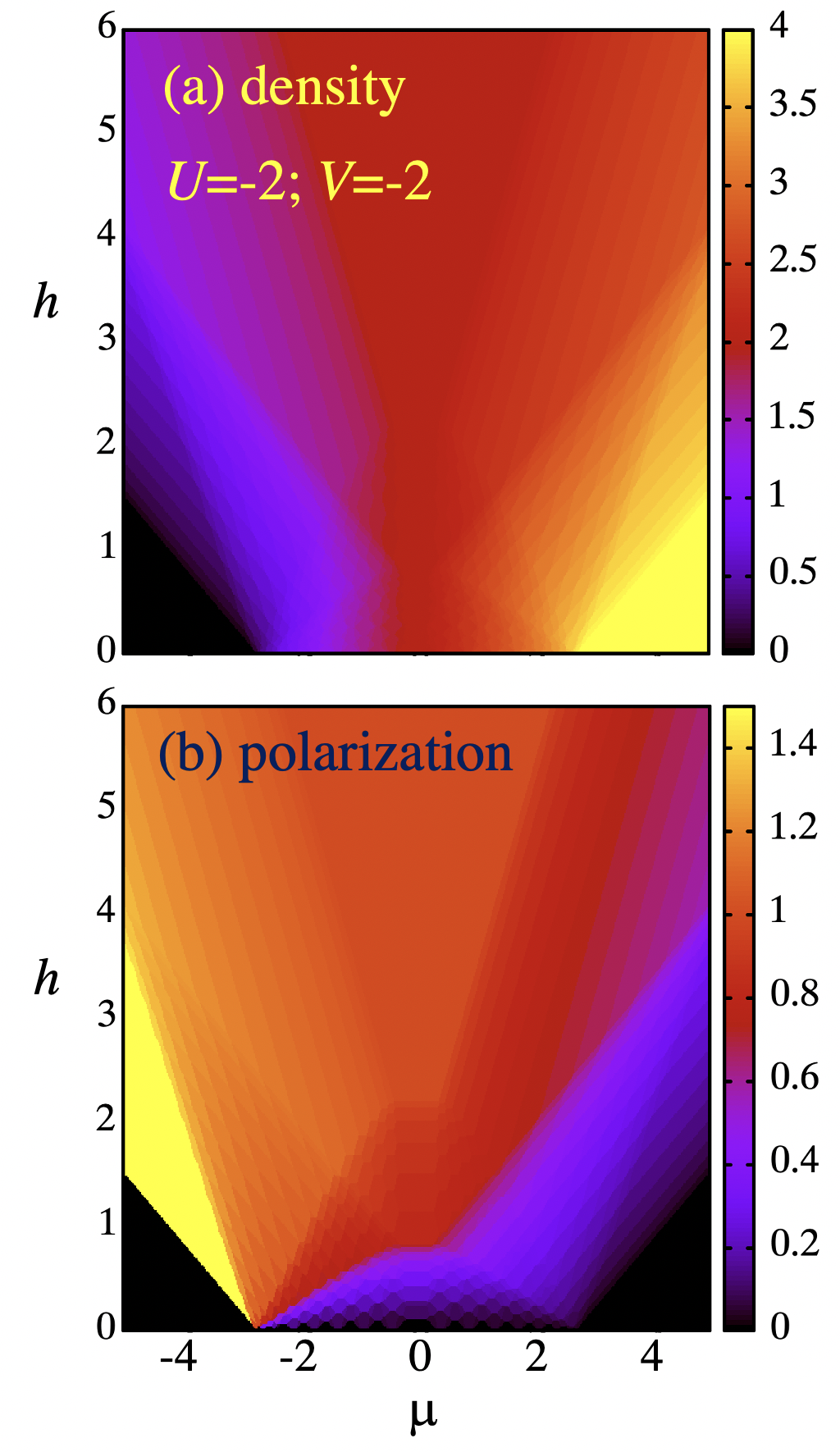}
    \caption{(a) Density and (b) polarization for a chain with $L=16$ sites as a function of magnetic field $h$ and chemical potential $\mu$, for $U=V=-2$.}
    \label{fig:phases_UV}
     \end{figure}
In this section we turn our attention to the case where the interaction $U$ is equal in magnitude to the single ion anisotropy $V$
($U$ = $V$). Following the same prescription used for the $SU(4)$ case, we have derived phase diagrams as well as correlation functions for the value $U = V=  -2$ as shown in Fig.\ref{fig:phases_UV}.

\begin{figure}
    \includegraphics[width=0.8\linewidth]{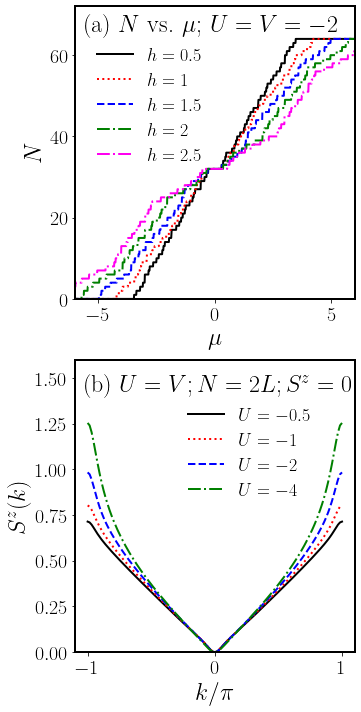}
    \caption{(a) Density as a function of chemical potential $n$ vs. $\mu$ for a chain with $L=16$ sites as a function of magnetic field $h$ and chemical potential $\mu$, for $U=V=-2$ and different values of the magnetic field $h$. (b) Longitudinal spin structure factor $S^z(k)$ for a chain of length $L=32$ at half-filling and for several values of $U=V$}
    \label{fig:spin}
     \end{figure}

Due to the single-ion anisotropy, the formation of local pairs with $S=2$ will be favored over pairs with $S=0$, incidentally energetically penalizing quartets as well. Therefore, at half-filling, zero polarization, and large $|U|$, we expect a Mott insulator with localized $S=2$ spins. In fact, it was already shown in Ref.\onlinecite{Tu2006} that in this regime, the model can be mapped to a generalized $S=2$ spin chain. As a consequence, the system will exhibit a single-particle charge gap for breaking a $S=2$ pair, and a Haldane spin gap. This is visualized in Fig.\ref{fig:spin}, where we show horizontal cuts of density as a function of the chemical potential, Fig.\ref{fig:phases_UV}(a), for several values of the magnetic field $h$. The incompressible phase can be clearly identified as a plateau at the particle-hole symmetric point (results for larger $U$ and larger $L$ --not shown here-- corroborate these findings).  

The presence of a magnetic field will break spin rotational invariance, further favoring pairs with large, positive $m=2,1$, such that at half-filling the system will behave as a partially polarized spin $S=2$ chain with a single-particle charge gap. Hence, near half-filling the single particle analogy is less intuitive. 

As we move away from half filling one would expect the system to resemble an $S=2$ {\it bosonic} $t$-$J$ chain, with holes moving in the spin background and with ``particles'' consisting of composite boson-like $S=2$ pairs. However, for small density and finite magnetic fields, kinetic energy may dominate and the single particle analogy is restored. For instance, at large negative $\mu$ and finite magnetic field $h$, the system starts filling up with $\alpha=3/2$ fermions first, as shown in Fig.\ref{fig:phases_UV}(b), and in similar fashion as the $SU(4)$ case. 

What remains to be seen is if the regions of partial occupation and partial polarization can support the same paired states as those found in the previous section. We have calculated correlations in the entire domain of densities and polarizations and found that the pair-pair correlations are completely featureless, decaying abruptly within two or three lattice spaces. This indicates that pairs are uncorrelated and behave as free-like hard-core bosons. Still, hard-core bosons can potentially condense, but in our case condensation is frustrated by the spin order, that dominates the physics. In Fig.\ref{fig:spin} we show the longitudinal spin-spin correlations along the $S^z$ projection, $\langle S^z_iS^z_j \rangle$, in momentum space, {\it i.e.} the static spin structure factor at half- filling ($N=2L$, $S^z_{Tot}=0$). Away from half-filling, $S^z(k)$ displays a singular peak at $k=k_F$ (instead of $2k_F$ because $S=2$ spins are composite particles formed by two fermions), as expected from a doped $t-J$ chain \cite{Moreno2011,Yang2022} (not shown).

\section{Conclusion}
To conclude, we have numerically obtained the phase diagrams of a generalized spin-3/2 Hubbard chain in terms of density and polarization as a function of magnetic field and chemical potential. By determining the partial occupation of the different spin channels we can predict the structure of the pairing order parameter. In agreement with previous work\cite{Szirmai2017}, in the $SU(4)$ sector of the Hamiltonian we observe a competition between ``quartetting'' and atomic density waves (ADW) for zero external field ($S^{z}$ = 0). Examining correlations in momentum space $k$ reveals that, at finite polarization, this phase evolves into a featureless Luttinger-liquid metal. As the magnetic field is further increased there is competition between the various ``quintet'' pairing channels. In the parameter regime studied in this work, we do not find evidence of trion or quartet long range order at finite polarizations. Instead, the system shows quasi-long-range order in the ``quintet'' channels with an order parameter with finite center of mass momentum, akin the FFLO states observed in spin-$1/2$ mixtures. Similar analysis was then extended to the non-$SU(4)$ sector for our spin-3/2 Hamiltonian ($U$ = $V$). At half-filling and zero polarization the system behaves as an $S$ = 2 Mott insulator (similar to an $S=2$ spin chain), confirming previous theoretical predictions \cite{Tu2006}. Clearly, this means that at half-filling, the system undergoes a transition from a spin gapped/charge gapless ADW phase to a spin gapped/charge gapped Mott insulating phase at some finite value of $V$. Away from half-filling the system  behaves as a {\it bosonic} $t-J$ chain with holes moving in a background of $S=2$ spins. Our numerical results expand on previous work by extending the characterization of the model to the full regime of densities and polarizations. Our approach to the problem in the grand canonical ensemble offers a new perspective and an intuitive interpretation of the phenomenology of spin-3/2 chains. 

\acknowledgements
The authors are grateful to the U.S. Department of Energy, Office of Basic Energy Sciences for financial support. This project was initiated under grant No.DE-SC0014407 and completed under grant No. DE-SC0022311.


%
\end{document}